Long title: Network analysis of Japanese global business using quasi-exhaustive micro-data for Japanese overseas subsidiaries

Short title: Network analysis of Japanese global business

Jean-Pascal Bassino[1,2,3] Pablo Jensen[1,4,6,7,9,10] and Matteo Morini[1,4,5,6,7,8,10#]

[1] Ecole Normale Supérieure de Lyon, France
[2] IAO (Institute of East Asian Studies, ENS Lyon and CNRS) UMR 5062, Lyon, France
[3] IER (Institute of Economic Research), Hitotsubashi University, Tokyo, Japan (visiting researcher)
[4] Université de Lyon, France
[5] Inria
[6] CNRS
[7] UCB Lyon 1
[8] Computer Science Lab, LIP UMR 5668, Lyon, France
[9] Physics Lab, CNRS UMR 5672, Lyon, France
[10] IXXI, Institute of Complex Systems, Lyon, France

# Corresponding author

Jean-Pascal Bassino contributed to: conceptualization, data curation, funding acquisition, investigation, methodology, resources, validation, writing–original draft preparation, and writing–review & editing. Pablo Jensen contributed to: conceptualization, funding acquisition, investigation, methodology, validation, and writing–review & editing. Matteo Morini contributed to: conceptualization, data curation, formal analysis, investigation, methodology, project administration, resources, software, validation, visualization, writing–original draft preparation, and writing–review & editing.

Abstract

Network analysis techniques remain rarely used for understanding international management strategies. Our paper highlights their value as research tool in this field of social science using a large set of micro-data (20,000) to investigate the presence of networks of subsidiaries overseas. The research question is the following: to what extent did/do global Japanese business networks mirror organizational models existing in Japan?
In particular, we would like to assess how much the links building such business networks are shaped by the structure of big-size industrial conglomerates of firms headquartered in Japan, also described as HK.
The major part of the academic community in the fields of management and industrial organization considers that formal links can be identified among firms belonging to HK. Miwa and Ramseyer (Miwa and Ramseyer 2002; Ramseyer 2006) challenge this claim and argue that the evidence supporting the existence of HK is weak.
So far, quantitative empirical investigation has been conducted exclusively using data for firms incorporated in Japan. Our study tests the Miwa-Ramseyer hypothesis (MRH) at the global level using information on the network of Japanese subsidiaries overseas. We identify linkages among Japanese subsidiaries overseas using an objective criterion: the subsidiaries share at least two Japanese co-investors (firms headquartered in Japan). The results obtained lead us to reject the MRH for the global dataset, as well as for subsets restricted to the two main regions/countries of destination of Japanese foreign investment: China (broadly defined as to include Hong Kong and Taiwan), and Southeast Asia. The results are robust to the weighting of the links, with different specifications, and are observed in most industrial sectors; the main exception is the automotive industry for which a straightforward explanation (unrelated to the MRH) exists. The global Japanese network became increasingly complex during the late 20th century as a consequence of increase in the number of Japanese subsidiaries overseas but the key features of the structure remained rather stable. We draw implications of these findings for academic research in international business and for professionals involved in corporate strategy.

# I. Introduction

The heuristic and explanatory power of network analysis techniques is widely acknowledged in various disciplines of social science (Scott, 1999; Padgett & Powell, 2012). They remain however rarely used for the empirical analysis of international management strategies, except in a few recent studies based on relatively small samples (e.g. Shi, Sun, Pinkham & Peng, 2014; for a review, see Hoang and Yi 2015). This remark applies also to academic research in management but also, surprisingly, to professional strategic consulting and business auditing activities. The results presented in the volume edited by David & Westerhuis (2014) demonstrate that a strong interest is emerging for studies with country-level historical perspective, including on Japan (Koibuchi & Okazaki 2014).

Our paper highlights the value of network analysis as a valuable research tool in international business with a study using a large sample of micro-data for the global network of Japanese subsidiaries overseas. The topic we investigate is related to an unsettled issue in Japanese business history that remains entirely relevant for analyzing present-day Japanese business strategies at home and abroad.

Specifically, we investigate if the structure of Japanese business networks is reminiscent of the corresponding horizontal conglomerates (HK). The majority of the academic community in the fields of management and industrial organization considers that the links between firms belonging to these HK can be identified through information on main-bank, cross-ownership, and transactions (e.g. Gerlach 1992; Aoki and Saxonhouse 2000). Miwa and Ramseyer (Miwa and Ramseyer 2002; Ramseyer 2006) challenge this claim, arguing the evidence supporting the hypothesis is weak. Alternatively, they interpret it as an ideological construct that was firstly devised by Japanese Marxists in the 1950s to be, later on, adopted by the Dodwell, a marketing company, and which was finally endorsed by non-Marxist scholars as well. As a concluding remark, it is worth stressing that quantitative empirical investigation have been conducted, up to now, exclusively using data for firms incorporated in Japan.

Our paper tests the Miwa-Ramseyer hypothesis (MRH) at the global level using information on Japanese subsidiaries overseas. The data are obtained from a nearly exhaustive global survey of Japanese overseas subsidiaries conducted by a private company, the Toyo Keizai Shinposha. Their dataset (henceforth TKZ) includes more than 20,000 firms in total, of which around 6,000 in the manufacturing sector, the one relevant to our analysis. The coverage is global and includes all recipient countries of Japanese foreign direct investment. The TKZ database reports information for wholly owned companies or joint ventures with local partners. Available information enables identifying Japanese and non-Japanese investors, and the shares owned by each firm. Membership of Japanese parent companies in one of the HK is defined on the basis of two indicators also supplied by TKZ: involvement regular meetings and equity ownership by firms identified core members of the HK. We use community detection techniques with different specifications and subsets of data in order to assess robustness of our results. Although with some caveats, the results obtained lead us to reject the MRH.

The remainder of the paper is organized as follows: section II offers an overview of the state of the art in business network analysis, and identifies some major gaps in the literature; section III describes the TKZ dataset; section IV describes the hypotheses under investigation; section V describes the strategy adopted to test the MRH and discusses the results; section VI summarizes the findings and elaborates on their implications for scholars and managers.

## II. Non-Japanese and Japanese business networks: state of the art and gaps in the literature

The major part of the studies on business networks remain focused on the analysis of relatively simple networks, either of centered on one particular firm, or using a small sample of observations.

What are the reasons for this limited development of complex network analysis in management, compared to other fields of social science, in particular economics or sociology? The three major intertwined explanations seems related to the strong focus typical of studies in management: (a) qualitative techniques (Harvard Business School type case studies), (b) intra-firm networks, and (c) directed networks (as opposed to undirected networks that are usually the more complex ones).

First, qualitative techniques enable analyzing extremely complex conditions, structures and strategies, but usually for one particular firm or, at best (from the viewpoint of network analysis), for a relatively small group of firms (e.g. Forsgren & Johanson 1992). In some minority cases, the scope is a little wider since the interactions of financial holdings and other investors are taken into account on the basis of information of equity holding.

The focus on intra-firm networks is a legacy of a tradition in management going back to Chandler (1962, 1977) that analyses the firm structural changes in response to the top management decisions meant to adjust corporate business strategies.

Finally, the focus on directed networks is also perfectly understandable: business networks are viewed essentially as centered on a group of top managers. The types of inter-firms networks considered are mostly hierarchical pyramids of firms linked by equity ownership relations with different tiers corresponding to subsidiaries of first, second, third, or $N$th rank. An interest for studies taking into account both strong and weak business ties using network analysis is however emerging recently (Kilkenny and Fuller-Love 2014).

In this context, the specialized sub-field of studies on Japanese business networks stands aside. The theory of the Japanese firm as a nexus of treaties formulated by Aoki (1984a, 1984b) emerged at the time when Japanese business networks were barely discussed (the book by Kono 1984 on the strategy and structure of Japanese enterprises, one of the most widely circulated in English around that period, does not mention business networks at all). Aoki's game-theoretical approach, although concentrated on intra-firm aspects, was very influential, as it provided an analytical framework applicable to undirected inter-firm networks. This solid conceptual bases enabled the development of an empirical stream of research that was later identified as keiretsu studies (in particular Gerlach 1992; Lincoln, Gerlach & Takahashi 1992; Lincoln, Gerlach & Ahmadjian 1996; Weinstein & Yafeh 1995; Aoki and Saxonhouse 2000; Nakamura 2002; Lincoln & Gerlach 2004). McGuire & Dow (2009) provides the most extensive recent survey of this stream of research.

The Japanese term keiretsu, which is usually translated as 'alignment' indicates that a firm has a set of preferential cooperations with another firm, generally bigger and in that case the relation is clearly hierarchical, or with a group of firms. The relation of the firm with this group can be either hierarchical, as aforementioned, or not hierarchical, in this latter case the firm is a member of an undirected network. In the case of hierarchical relations between firms, the structure is the same as in vertically organised business groups that exist in all regions of the world. This pattern is described in

Japanese business studies as vertical keiretsu (VK), while the undirected network is described as horizontal keiretsu (HK). It should be noted, however, that there are overlaps between VK and HK. Specialized manufacturing groups sur as Toyota and Mitsubishi are vertically organized and, at the same time, they are part of an HK (Mitsui and Mitsubishi keiretsu, respectively.)

The development of the Keiretsu studies led, among other consequences, to revive the interest in Japanese business history studies investigating the strategies and structures of prewar groups owned by kinship networks, i.e. zaibatsu such as Mitsui, Mitsubishi, Sumitomo, and Yasuda, dissolved in 1946 upon request of the U.S. occupation authorities. An obvious issue was assessing the strength of post-war links between companies that belonged to these pre-war groups and that morphed in the 1950s into HK type conglomerates whose membership was somehow different from the Zaibatsu's (see also Ramseyer & Miwa (2007) for a discussion on the dissolution of pre-war zaibatsu). Ironically, the expansion of the literature on HK accelerated in the 1990s and 2000s, precisely at the time when the ties were becoming increasingly informal and weak. What is more, the description of the structure of HK became obsolete as a consequence of waves of mega-mergers of the major Japanese banks that took place in 2006. This led some of the key figures in the Keiretsu studies to reflect on their demise ("why they are gone?") and the future of Japanese business groups (e.g. Lincoln & Shimotani (2009); see also McGuire & Dow (2009) for a similar discussion).

Looking at studies concerning Japanese business network overseas, we can observe that a number of papers are explicitly referring to keiretsu membership for investigating various issues such investment overseas (Belderbos and Sleuwaegen 1996), as spatial location decision and agglomeration effects (e.g. Belderbos, & Sleuwaegen 2002, Yamashita, N., Matsuura & Nakajima 2014). Zhang, M. M. (2015). A few papers also discuss the importance of keiretsu in shaping Japanese business networks overseas, in some case using the large datasets such as the TKZ database (e.g. Zhang 2015). However, to the best of our knowledge, there has been no attempt so far to use standard network analysis techniques to unravel the structure of Japanese networks overseas and our study aims at filling this gap.

## III. Dataset and descriptive statistics

The TKZ dataset we use actually is composed by two datasets, both produced by the Toyo Keizai Shinposha, a private company whose denomination in English is Oriental Economist (http://corp.toyokeizai.net/en/). One of the attractive features of this database is that it has not been constructed by the Japanese government or a not-for-profit semi-public body, but rather by a private company: henceforth, their quality and accuracy was meant to generate a positive return on investment. Thus, that these volumes are rather best sellers than confidential publications is indicative of such quality and of the trust the public had in the information supplied. The surveys have been repeated yearly since more than 50 years on the basis of voluntary participation.

The first database reports micro-data resulting from a yearly survey administered in 2005 to Japanese subsidiaries based overseas (a sizeable 20.700), circulated in the 2006 TKZ edition that we use for reasons explained below. The firms respond on a voluntary basis, and some piece of information is sporadically missing in the returned forms, collected and processed by Toyo Keizai. The sampling rate is not disclosed by the Toyo Keizai, but the consensus is that coverage is extremely high because the respondents are not expected to report confidential information. The list of data requested is limited to the denomination, address, industrial sector, paid-up capital, name and share of each Japanese equity owner, and share of local investors, when joint ventures are established with local foreign partners; notably, the respondents are not required to disclose the identity of the local partners. Toyo Keizai processes the information as to include a unique code for each subsidiary and, more importantly, for equity owners (companies headquartered in Japan; no code for local investor). The equity owners univocal coding system aptly implemented by Toyo Keizai prevents any risk of confusion due to mislabelling: the local subsidiaries managers responding to the survey are held back from using possibly inconsistent textual denominations. Since we had the chance to access the electronic version of the 2006 database (more often than not, the printed version is used), there is no risk of error or omission (at least not by our research team).

We process the information available in the database to identify conglomerates of Japanese overseas businesses, by defining a quantitative and objective criterion: two Japanese investors (firms headquartered in Japan) are considered to be linked (as nodes connected by an edge on the graph) if they co-invest in, i.e. they co-own, one or more overseas subsidiaries.

These co-ownership relations are of utmost relevance since they are <u>measurable</u> in stark contrast with relations between firms only of informal cooperation and/or repeated transactions, without any equity ownership tie.

As side note, a subsidiary may be involved in more than one business network. Thus, such shared subsidiaries heuristically play the role of "bridges" between two networks, therefore they contribute to the network cohesion. For the interested reader, such links are related to the concept of *weak ties* (Granovetter 1973).

Furthermore, the advantage, among others, of the TKZ dataset is to include information on mid-size groups. However, after 2006, the mega-mergers of some Japanese banks disrupted such fine scale structure and for this reason we focus our analysis on the pre-2006 networks. The implications of our results on pre-2006 networks for analyzing present day conditions are discussed in section VI.

The second Toyo Keizai database we use, much smaller, has been obtained from the last issue (published in 2000) of the Toyo Keizai "Keiretsu Survey" (a distinct product from the Toyo Keizai yearbook on industrial groups, published yearly). It provides very valuable information on the 6 big HK: Mitsui, Mitsubishi, Sumitomo, Sanwa, Fuyo, and Ikkan, that can be considered as the indirect heirs of kinship-based zaibatsu ("financial cliques") such as Mitsubishi, Mitsui, and Sumitomo, that were dissolved in 1947. Relying on information obtained from Toyo Keizai (2000), we retain two criteria for membership in one of the 6 HK: (i) the firm is a member of one of the 6 Chief Executive Officers "clubs" (one per HK) meeting on a weekly/monthly basis; (ii.) the firm is among the top 50 companies by share of equity ownership of companies members of one of the 6 "clubs". To the best of our knowledge, the Toyo Keizai did not officially explain why the publication Keiretsu survey has been discontinued, the 2000 issue being the last one. Two alternative interpretations can be considered. The first one, which can be considered consistent with the Miwa-Ramseyer view, is that the *Toyo Keizai* finally acknowledged the fallacy in the Keiretsu existence nowadays. The second one, which we tend to favor, is that drastic changes in the organizational structure of Japanese business networks (in Japan, and - presumably and consequently - abroad) resulted in a dying out of the preferential links and cooperation networks that had been identified earlier. The information reported in the Keiretsu volume was therefore becoming less relevant and, at any rate, redundant with the one provided in the separate Toyo Keizai volume on business groups, more focused on the concept of vertical keiretsu. In particular, as aforementioned, the 2006 mega mergers of a number of Japanese banks rendered less and less relevant keeping a list of firm memberships in a "club".

Nevertheless, the 2000 database provided by the latest issue of the "Keiretsu Survey" is valuable to our study as it provided us with the backbone underlying the keiretsu web we test our hypothesis against.

Thus, summing up, the two dataset together concur to build a coherent perspective: from the first dataset we extract with network analysis tools, described in Sec. V, the granularity of the Japanese business network, i.e., its communities structure; on the other hand, we compare this empirical evidence of communities against the "Keiretsu Survey" dataset to infer if such communities mirror the HK organization.

## III.1 Worldwide waves of investment

Japanese foreign direct investment (FDI) did not took place in parallel in all regions of the world, but rather in successive waves. Japanese investors were first attracted in the 1960s and 1970s by the comparative advantages of ASEAN countries, Hong Kong and Taiwan, and the possibility to gain access to these emerging markets. From the 1970s, and especially during the 1980s, North America and Europe also became important destinations. Finally, the gradual opening of the People's Republic of China to international trade and foreign investors in the 1980s resulted in a reorientation of Japanese foreign direct investment that accelerated in the 2000s; China became the main target country in terms of flows, and - after some time lag - also in terms of stock. Evidence from the TKZ is presented in **Figure 1**.

**Figure 1. Japanese co-investment overseas (number of firms in 1965, 1975, 1985, 1995, and 2005; log scale)**

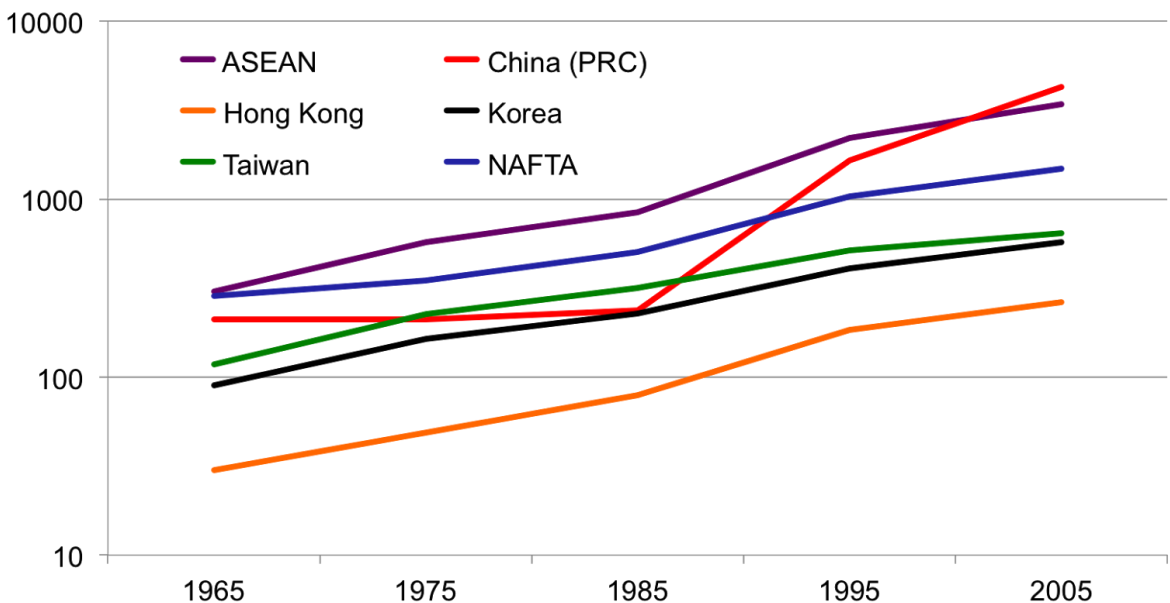

Note: The 10 countries of the ASEAN (Association of Southeast Asian Nations) are: Brunei, Cambodia, Indonesia, Laos, Malaysia, Myanmar, the Philippines, Singapore, Thailand, and Vietnam.

## III.2 Heterogeneity in ownership and prevalence of manufacturing

The ratio of TKZ subsidiaries which are included and used in reconstructing the Japanese investors business networks vary manifestly by country. This is both because of the overall quota of manufacturing enterprises (very low in the EU, high in ASEAN countries, cfr. **Figure 2**), and the ratio of single owned subsidiaries (very high in the EU, lower in ASEAN countries, **cfr. Tables 3.a, 3.b**, below).

Figure 2. % of Japanese FDI in USD (source: TKZ), by macroarea, all sectors vs. manufacturing only

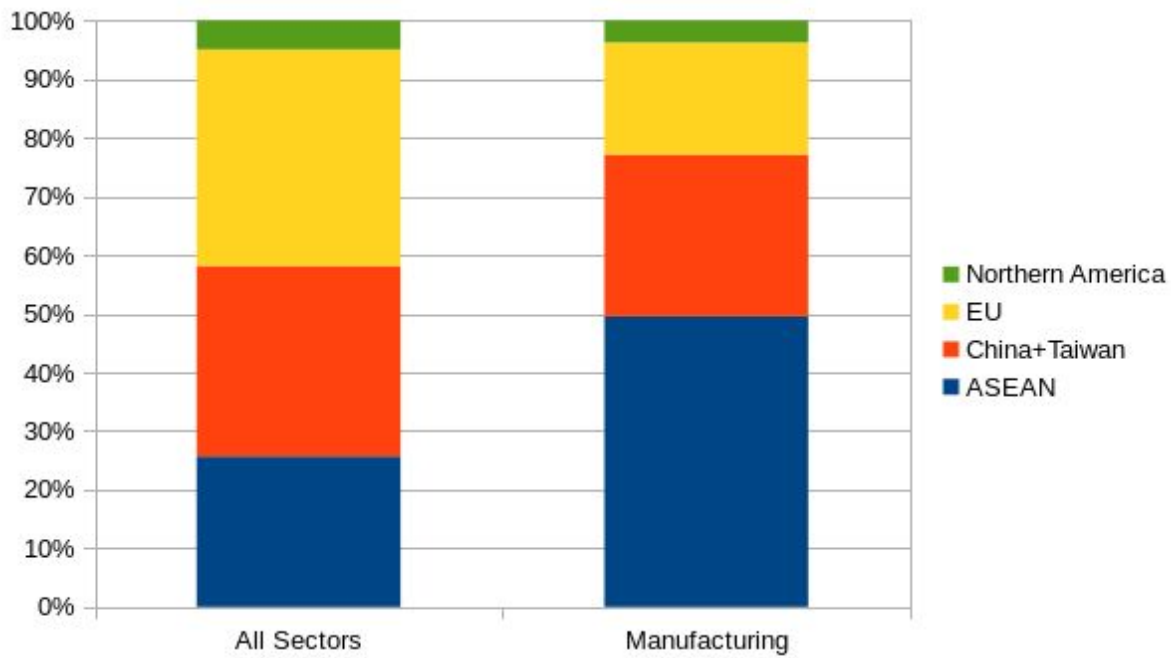

As mentioned previously, non-manufacturing businesses are not included. This is ostensibly visible in the most extreme cases, the Netherlands and Indonesia respectively (cfr. **Table 1**).

Table 1. Frequencies of firms by country: full TKZ dataset (TOT), firms included (INCL), % difference; first 20 countries sorted by decreasing TOT

| Country | TOT | INCL | DIFF% |
| --- | --- | --- | --- |
| China | 4.414 | 3.497 | -21% |
| USA | 3.419 | 1.194 | -65% |
| Thailand | 1.534 | 1.116 | -27% |
| Hong Kong | 1.109 | 254 | -77% |
| Singapore | 1.036 | 276 | -73% |
| Taiwan | 907 | 520 | -43% |
| United Kingdom | 806 | 188 | -77% |
| Malaysia | 774 | 516 | -33% |
| Korea | 682 | 453 | -34% |
| Indonesia | 673 | 571 | -15% |
| Germany | 637 | 126 | -80% |
| Philippines | 453 | 274 | -40% |
| Australia | 402 | 69 | -83% |
| France | 380 | 123 | -68% |
| Netherlands | 358 | 56 | **-84%** |
| Canada | 266 | 91 | -66% |
| Brazil | 261 | 137 | -48% |
| Viet Nam | 250 | 200 | -20% |
| Mexico | 219 | 124 | -43% |
| India | 199 | 157 | -21% |
| *(61 more countries omitted)* | (...) | (...) | (...) |
| **TOT** | **20.488** | **10.481** | **-49%** |

## III.3 Heterogeneity in subsidiaries size and capitalization

As it will become noticeable in the analysis, geographical and sectoral peculiarities are evident in the dataset, and play a relevant role in shaping the topological features of networks built from disaggregated data. In **Tables 2.a, 2.b**, descriptive data are shown over both dimensions.

Table 2.a. Average number of employees (NEMPL) and capitalization (K), disaggregated by macroarea

| Area | NUMEMPL | K |
|---|---|---|
| ASEAN | 518 | 21.980 |
| China and Taiwan | 366 | 12.204 |
| EU | 331 | 20.191 |
| Northern America | 401 | 37.900 |

Table 2.b. Average number of employees (NEMPL) and capitalization (K), disaggregated by industrial sector

| Industrial sector | Code | NUMEMPL | K |
|---|---|---|---|
| Food | 600 | 346 | 24.781 |
| Textiles | 700 | 380 | 5.117 |
| Wood and Furniture | 800 | 323 | 12.252 |
| Pulp and Paper | 900 | 246 | 31.144 |
| Publishing and Printing | 1000 | 159 | 283 |
| Chemicals and Pharma | 1100 | 167 | 18.894 |
| Petroleum and Coal Products (incl. Plastics) | 1200 | 55 | 66.326 |
| Rubber and Leather Products | 1300 | 258 | 6.185 |
| Refractories and Glass | 1400 | 477 | 6.859 |
| Steel | 1500 | 313 | 140.409 |
| Non-ferrous Metals | 1600 | 545 | 37.619 |
| Metal Products | 1700 | 218 | 11.905 |
| Machinery | 1800 | 236 | 11.244 |
| Electric and Electronic Machinery and Devices | 1900 | 832 | 27.436 |
| Transport Machinery and Shipbuilding | 2000 | 308 | 28.089 |
| Automobiles and Parts | 2100 | 563 | 31.848 |
| Precision Machinery | 2200 | 541 | 11.677 |
| Miscellaneous Manufacturing | 2300 | 253 | 4.531 |

## III.4 Co-investments

Globally, every subsidiary is owned by approximately one and a half (1.48) investors, unevenly split between Japanese (1.31) and local (0.17). In **Table 3.a**, this information is disaggregated by geographical macro-area; the European Union, where less than one fifth (17%) of the subsidiaries are owned by two or more investors, appears to be the area with the least average number of investors, closely followed by Northern America.

**Table 3.a. Co-investors per subsidiary, disaggregated by geographical area**

| Area | Avg | 1 | 2 | 3 | 4 | 5+ | (%) 2+ |
|---|---|---|---|---|---|---|---|
| ASEAN | 1.66 | 1.530 | 553 | 259 | 99 | 58 | 39% |
| China and Taiwan | 1.52 | 2.190 | 774 | 281 | 86 | 37 | 35% |
| EU | 1.21 | 648 | 102 | 20 | 6 | 1 | 17% |
| Northern America | 1.28 | 967 | 189 | 55 | 8 | 3 | 21% |

**Table 3.b** shows a very high ratio of extensively co-owned firms for a few sectors (e.g. textile, where more than half of the subsidiaries are co-owned by two or more investors), whereas single-ownership appears to be the norm for others (e.g. precision machinery, including one eighth of co-owned subsidiaries only).

**Table 3.b. Co-investors per subsidiary, disaggregated by industrial sector**

| Industrial sector | Code | Average | 1 | 2 | 3 | 4 | 5+ | (%) 2+ |
|---|---|---|---|---|---|---|---|---|
| Food | 600 | 1.59 | 306 | 101 | 48 | 18 | 7 | 36% |
| Textiles | 700 | 1.83 | 252 | 151 | 74 | 30 | 7 | 51% |
| Wood and Furniture | 800 | 1.57 | 52 | 14 | 5 | 1 | 4 | 32% |
| Pulp and Paper | 900 | 1.81 | 47 | 34 | 15 | 1 | 3 | 53% |
| Publishing and Printing | 1000 | 1.11 | 50 | 4 | 1 | 0 | 0 | 9% |
| Chemicals and Pharma | 1100 | 1.52 | 995 | 343 | 120 | 40 | 18 | 34% |
| Petroleum and Coal Products (incl. Plastics) | 1200 | 2.05 | 9 | 3 | 4 | 3 | 0 | 53% |
| Rubber and Leather Products | 1300 | 1.41 | 177 | 42 | 24 | 4 | 0 | 28% |
| Refractories and Glass | 1400 | 1.42 | 174 | 53 | 17 | 5 | 1 | 30% |
| Steel | 1500 | 2.27 | 74 | 50 | 39 | 15 | 16 | 62% |
| Non-ferrous Metals | 1600 | 1.59 | 155 | 43 | 21 | 10 | 5 | 34% |
| Metal Products | 1700 | 1.63 | 261 | 112 | 40 | 22 | 5 | 41% |
| Machinery | 1800 | 1.38 | 803 | 194 | 55 | 16 | 13 | 26% |
| Electric and Electronic Machinery and Devices | 1900 | 1.26 | 1518 | 272 | 68 | 14 | 8 | 19% |
| Transport Machinery and Shipbuilding | 2000 | 1.54 | 43 | 20 | 8 | 1 | 0 | 40% |
| Automobiles and Parts | 2100 | 1.64 | 767 | 339 | 135 | 42 | 20 | 41% |
| Precision Machinery | 2200 | 1.15 | 224 | 30 | 2 | 0 | 1 | 13% |
| Miscellaneous Manufacturing | 2300 | 1.23 | 279 | 41 | 10 | 3 | 1 | 16% |

# IV. Hypotheses

Before starting our analysis, we would like to briefly sketch the underlying hypotheses to our approach that guided us and which were, as we explain in the following, mostly driven by the available information provided by the dataset, as in the case of the geographical distribution of the Japanese foreign investments, and by sensible heuristical observations on the data's nature.

**Hypothesis 1:** considering the manufacturing firms included in the TKZ dataset in all countries for 2005, we reject the strong form of the Miwa-Ramseyer hypothesis.

With regard to the Miwa-Ramseyer Hypothesis (MRH), we adopt an agnostic view. Indeed, we accept their claim that the empirical evidence supporting the existence of HK is rather weak.
These business groups would be particularly difficult to identify should the affiliations be informal, implying that they would not require any kind of binding and irreversible commitment. Moreover, anecdotal evidence indicates that a number of firms that were identified as informal members of one HK gradually shifted to an equally informal affiliation with another HK. What is more, a number of firms loosely related to a HK eventually moved to a position of dual affiliation.
Therefore, it is not surprising that the evidence obtained using panel data analysis or similar econometric techniques could be disappointing. However, we do not reject the possibility that a nexus of bilateral or multilateral treaties and repeated transactions between firms, as well as information exchanges, involvement in joint R&D projects, and cooperation in joint ventures at home and abroad could result in the formation of an indirect business network involving tightly knit clusters of firms.

Furthermore, one of the main advantages of network analysis in this context is to be consistent with the possibility of informal affiliation and/or changes in affiliation. Our first hypothesis is that Japanese business networks overseas tend to replicate familiar structures already in place in Japan when HK had very few foreign subsidiaries, that is in the 1960s and 1970s. Accordingly, what we would describe as a strong form of MRH ("[...] at root, the keiretsu do not exist." in Miwa and Ramseyer prose), can be rejected.

**Hypothesis 2:** the structure of Japanese business networks overseas is becoming gradually more complex but the identified key-players remained essentially the same ones during the period 1975-2005.

To what extent did the structure of Japanese business networks overseas evolved over time? The firms setting up foreign subsidiaries in the 1960s and 1970s have been overwhelmingly the biggest players in their industry, and in their respective HK, if we believe the proponents of the strong form of the horizontal keiretsu hypothesis (HKH). It is only with the dramatic increase in volume of foreign direct investment that mid-size firms (or Japanese-based firms situated in the periphery of the HK, according to the the strong form of HKH proponents) became present as parent companies of foreign subsidiaries. Since our measure of involvement in a Japanese business network overseas is defined precisely as the co-investment in foreign subsidiaries with other Japanese investors, we expect to find a high level of stability in the structure observed using information successive benchmark years. We selected four benchmark years with a 10-year interval between them: 1975, 1985, 1995, and 2005.

**Hypothesis 3:** consistent results are expected with or without weighting.

Weighted networks may provide an alternative insight as to the business links used to identify groups of firms on the network. Instead of mere binary yes/no relationships, a metric can be put in place to define the links between any two investors, in terms of a continuum measuring the strength of their economic ties. For instance, when considering the shared set of subsidiaries common to two investors, as an example of the weight, both the capitalization and the owned share must be taken into consideration. See section V.3 for details.

**Hypothesis 4:** Similar results (rejection of the MRH) are expected, even when considered separately, for three out of the four main (in terms of destination of Japanese foreign investment) world macro-regions: ASEAN, China (including Hong Kong and Taiwan), and North America. Not enough observations are available for Europe, on the other hand, in order to present conclusive results.

As observed in Section III.1, foreign investments for Japanese firms historically proceeded by successive waves, first hitting ASEAN countries, to then spread to North America and Europe and, finally, to China. However, if the architecture of the Japanese business networks in these different regions of the world was determined by characteristics of the links between Japanese parent companies, we would expect to find similar results.

**Hypothesis 5:** When considering the peculiar structure of some industrial sectors, similar results may not be observed in all of them.

The investors and subsidiaries involved in the Japanese business networks overseas specialised in different lines of business. That is also the case of the big conglomerates (HK). However, the major part of the co-investments are likely to associate firms of the same industrial sectors, in particular in the manufacturing sector. Depending on the number of potential partners and the advantages derived from co-investments in foreign subsidiaries in terms of information sharing and risk mitigation, it is conceivable that firms that would be normally competitors may decide to cooperate in order to penetrate a foreign market. It is therefore plausible that such a specific pattern, contradicting the general trend, is observed in some sectors, e.g. automotive, food and textile.

# V. Methodology

## V.1 Community detection on inferred business network

We test the MRH using a standard network community detection technique (Blondel et al. 2008) on the reconstructed co-investment graph described in Sec. III, matching the communities each node (firm) is assigned to, and information on real-world business conglomerates (keiretsu), in order to evaluate to what extent the communities detected in the networks correspond to communities defined by at least one of the criteria of membership of one of the 6 big HK obtained from Toyo Keizai (2000).

The results are tested against networks which are equivalent, degree-wise, but with randomized structure (Configuration Models, Newman 2003, as null hypotheses). Absence of significant correlation between communities and keiretsu is consistently shown in the latter case.

The network is built using Japanese investors appear as nodes, which are linked if they share investments in at least N subsidiaries. Figure 3 offers an insight of the role of subsidiaries in structuring clusters of Japanese investors.

**Figure 3. Heterogeneous network, including Japanese investors and overseas subsidiaries, ASEAN countries only**

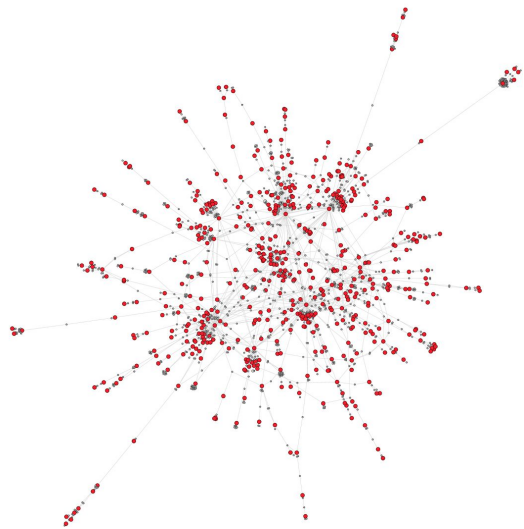

*Note: red and grey nodes represent, respectievly, Japanese investors and subsidiaries; sizes are arbitrary and equal within each category, in order to better reveal the underlying network structure.*

*Zoomable version available at http://perso.ens-lyon.fr/matteo.morini/jpbusnet/Het_ic.pdf ; industrial codes (see Table 3.b) are readable as labels in the online version.*

Both parent companies in Japan and subsidiaries overseas are reliably identified by unambiguous Toyo Keizai codes. Data on foreign local partners is also available, but their identification is hardly consistent due to the absence of unique identifiers; many generic "anonymous" nodes also appear. At any rate, their contribution to the network structure appears negligible (cfr. III.4), which led us to finally omit this information in order to avoid unpredictable bias.

The main purpose of representing TKZ microdata as a network of co-investments is to look for "more densely connected" sets of firms which are, by definition, communities. Firms tend to cluster together in the network structure *if* there are prevalent and privileged economic links between them. If we observe a non-random distribution (systematic overrepresentation) of firms known to belong to a Keiretsu across communities, we can conclude that the economic structure revealed by the network topology is driven by Keiretsu-type inter-firm bonds.

Our hypothesis can be tested by contrasting two different characteristics of every firm: on the one hand, a firm can be a member of either one of the Big 6 Keiretsu groups; on the other, it belongs to one of the communities derived from the network structure. We seek to verify the independence (or lack thereof) between the economic network, as resulting from the community detection method described above, and the Keiretsu structure.

Figure 4 displays the bivariate joint frequencies of firms on both Keiretsu and community categories as a heatmap, for the global dataset, including manufacturing firms worldwide. Hints of a non-random distribution of firms across the two categories, to be validated statistically (see sec. V.2), are visible in the image. Mitsubishi- and Sumitomo-bound investors, for instance, are ostensibly concentrated in two communities.

### Figure 4. Bivariate joint frequencies of Japanese investors; worldwide, manufacturing sector

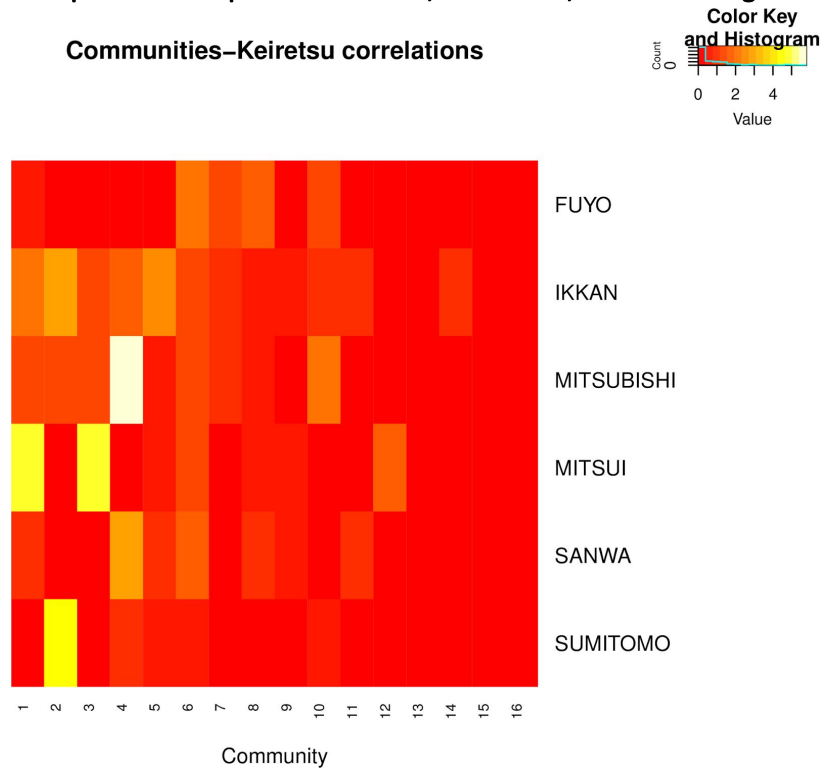

## V.2 Independence test

An associative measure must be put to use; since we are dealing with categorical data, the Pearson Chi-Square test, including some case-specific safeguards, have been deemed as appropriate. It has been applied to the resulting two way table, to assess whether there is any interdependence between the two attributes (this being the alternative hypothesis $H_a$, if the null hypothesis of independence $H_0$ can be rejected).

Moving on to assess the **test validity, we observe that** the population is fairly large, and lends itself to an in-depth analysis of subsets of interest. However, in a few specific instances (e.g. "Europe" macro-region, "Food industry" industrial sector) the number of selected observations is barely adequate. For the sake of robustness, a double line of defense has been put into place: first, a key characteristic of the Louvain community detection method has been leveraged in order to increase the number of observations: being based on a stochastic algorithm, repeated runs (1.000 iterations, in this case) return slightly different partitions, which can be accumulated into a richer dataset (and averaged out, to keep from artificially inflating the sample size, biasing the test); second, the Chi-Square p-values have been computed by Monte Carlo simulation (Tate and Hyer 1973, Bradley and Cutcomb 1977).

## V.3 Business ties as weighted links

The strength of business ties, in weighted networks, is measured as Cosine Similarity (Salton and McGill 1983). CS compares the distribution of the capital invested by pairs of co-owners into subsidiaries: a perfect match (e.g. $K_{a,1} = K_{b,1}$; $K_{a,2} = K_{b,2}$; ... $K_{a,n} = K_{b,n}$, where $K_{PC,S}$ is the capital K invested by the parent company PC in subsidiary S) corresponds to a CS = 1; as the allocation choices diverge, CS approaches zero. Technically, it is a measure of the angle between the two vectors, and its purpose is to offer a proportional representation of the connection between investing firms going beyond the binary idea of connected vs. unconnected.

# VI. Results and discussion

Timewise, it appears that, after the first, sparsely populated 1975 snapshot, a period of strong correlation (rejection of the MRH), ensues, encompassing the following decades (1985 and 1995 snapshots). In 2005, right before the mega-mergers occur, evidence starts to wane.

Geographically, an indisputable difference is observable between macro-areas with medium to strong significance (Asean countries, China and Taiwan, Northern America) and one area where there is no evidence for the persistence of HK structures at all (Europe).

When a disaggregated analysis is performed by industrial activity, the only, albeit extremely sizeable, sector with unambiguously strong correlation is "chemical"; "textile" does not offer a strong enough evidence for the existence of HK (there appears to be a very weak correlation for the unweighted network case); the independence hypothesis cannot be rejected for "food" either, since the small number of observations keeps us from achieving robust and conclusive results. Automotive yields ambiguous results (correlated when weighted, uncorrelated when unweighted); a tentative explanation, which would require a more in-depth analysis out of the scope of this paper, may lie in the presence of numerous "smaller" partners, playing a minor role. Links implying these partners are weaker, and a more clear-cut HK structure appears when considering the most economically important subsidiaries only, connected by stronger links. The root of this discrepancy might be traced back in the bulk of smaller partners whose importance is correctly rescaled through the links' weight, thus evidencing the correlation with the HK.

Table 4: Hypotheses test results, unweighted (left) and weighted (right) columns

|  | UNWEIGHTED | | | WEIGHTED | | |
| --- | --- | --- | --- | --- | --- | --- |
|  | *MRH Rejected?* | *ChiSq* | *p-val* | *MRH Rejected?* | *ChiSq* | *p-val* |
| Overall | YES * | 72.15 | 0.013 | NO | 56.18 | 0.080 |
| 1975 | YES * | 45.65 | 0.039 | NO | 56.10 | 0.140 |
| 1985 | YES *** | 61.07 | 0.001 | YES *** | 96.70 | 0.000 |
| 1995 | YES *** | 92.77 | 0.000 | YES *** | 91.64 | 0.000 |
| 2005 | YES * | 72.15 | 0.013 | NO | 56.18 | 0.080 |
| ASEAN | YES * | 52.68 | 0.030 | YES *** | 67.83 | 0.000 |
| China and Taiwan | YES *** | 77.51 | 0.000 | YES *** | 57.37 | 0.000 |
| Northern America | YES * | 78.96 | 0.011 | YES *** | 77.50 | 0.000 |
| EU | NO | 41.55 | 0.254 | NO | 49.11 | 0.150 |
| Automobiles and Parts | NO | 70.39 | 0.982 | YES *** | 91.10 | 0.000 |
| Food | NO | 45.67 | 0.320 | NO | 45.67 | 0.310 |
| Chemicals and Pharma | YES *** | 73.36 | 0.000 | YES *** | 63.43 | 0.000 |
| Textiles | NO | 23.32 | 0.080 | NO | 28.77 | 1.000 |

Note: *** p < .001; ** p < .01; * p < .05

In order to assess the validity of our findings, the results for every hypothesis tested have been contrasted to an alternative network Configuration Model (Newman 2003), where the network structure is destroyed, while preserving the degree for single nodes, through a random rewiring process. Intuitively, this procedure is equivalent to blindly creating economic partnerships. In every single instance, any hint of significance disappears completely, showing p-values very close to 1.

Figure 5. Worldwide Japanese investors network: business ties and Big-6 membership.
Highlighted: Mitsubishi and Mitsui clusters

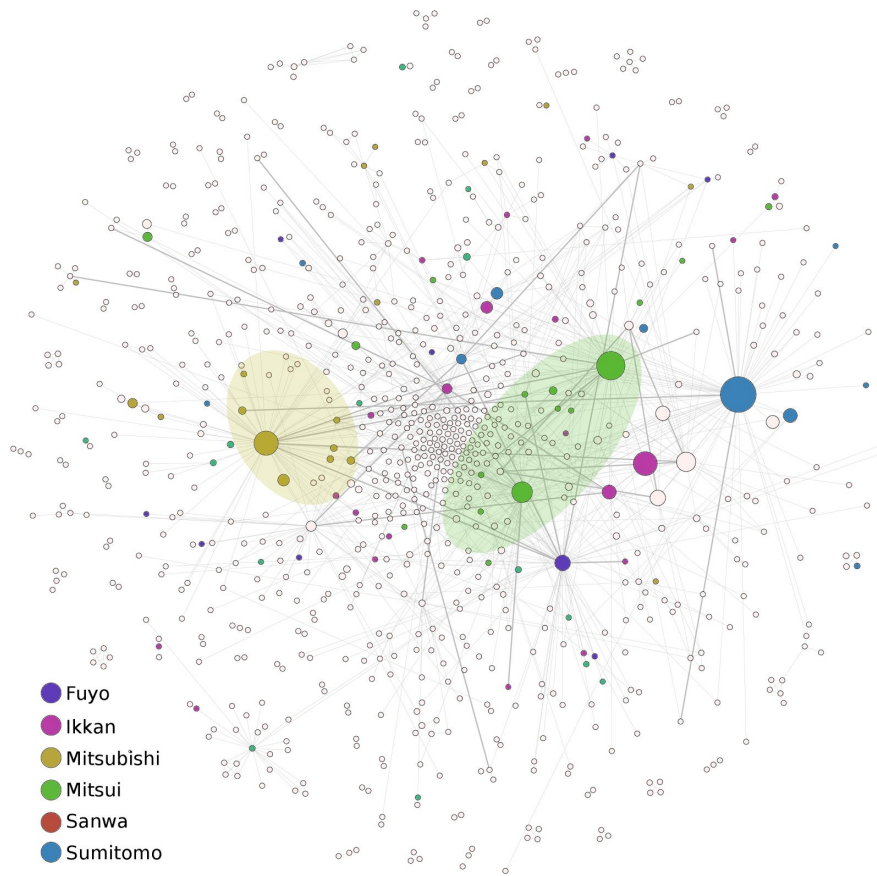

## VII. Conclusions

In this work, we have presented an analysis of the TKZ dataset, using community detection tools, that led us to reject the MRH in the strong form. Indeed, we were able to display quantitative evidence that the communities embedded in Japanese business networks strongly correlate with the HK structure described in the 2000 "Keiretsu Survey" dataset.

To give a brief summarizing overview, as a first step, we extracted from the 2006 TKZ dataset a graph of co-ownership, so that two firms are linked in our approach if they both invest in an overseas subsidiary. Through community detection algorithms, the taxonomy of high-density clusters emerged from this graph so that each firm was classified into one cluster, as described in Sec. VI. This classification, purely arising from explicit business ties (the co-ownership), was then compared to the Keiretsu one, described in the 2000 TKZ dataset.

Our results, summarized in Table 4, strongly point to a clear correlation between the intrinsic network organization and the HK, albeit with some fluctuations when one considers more specific subsets, eventually flawed by the lack of statistics, as for the investments in Europe and the food industry.

Another interesting point unveiled by the analysis is, in the automotive sector, the discrepancy between the clear correlation shown by the weighted network with respect to the unweighted configuration.

Finally, to test the soundness of our findings, we provided, as a comparison, a null model by shuffling the links and destroying the existing correlations with the Configuration Model. This test, disrupting any network structure, leads to p values near to 1 and, thus, further proves that the Japanese business network bears a strong intrinsic mark in its organization.